\documentclass[a4paper,11pt]{article}
\pdfoutput=1

\usepackage{jcappub}

\usepackage[T1]{fontenc}
\usepackage{pstricks}
\usepackage{color}
\usepackage{graphicx}
\usepackage{multirow}
\usepackage{slashed}
\usepackage{subfigure}
\usepackage{hyperref}
\usepackage{amsmath}
\usepackage{amssymb}
\usepackage{amsfonts}
\usepackage[utf8]{inputenc}
\usepackage{feynmp}
\usepackage{feynmp-auto}
\title{The Dark Side of the Littlest Seesaw: freeze-in, the two right-handed neutrino portal
and leptogenesis-friendly fimpzillas}

\author[a]{Marco~Chianese}
\author[a]{and Stephen~F.~King}

\affiliation[a]{Department of Physics and Astronomy, University of Southampton, SO17 1BJ Southampton, United Kingdom}

\emailAdd{ma.chianese@gmail.com}
\emailAdd{king@soton.ac.uk}

\abstract{
We propose a minimal model to simultaneously account for a realistic neutrino spectrum through a type-I seesaw mechanism and a viable dark matter relic density. The model is an extension of the Littlest Seesaw model in which the two right-handed neutrinos of the model are coupled to a $Z_2$-odd dark sector via right-handed neutrino portal couplings. In this model, a highly constrained and direct link between dark matter and neutrino physics is achieved by considering the freeze-in production mechanism of dark matter. We show that the neutrino Yukawa couplings which describe neutrino mass and mixing may also play a dominant role in the dark matter production. We investigate the allowed regions in the parameter space of the model that provide the correct neutrino masses and mixing and simultaneously give the correct dark matter relic abundance. In certain cases the right-handed neutrino mass may be arbitrarily large, for example in the range $10^{10}-10^{11}$ GeV required for vanilla leptogenesis, with a successful relic density arising from frozen-in dark matter particles with masses around this scale, which we refer to as ``fimpzillas''.}

\begin{document} 
\maketitle
\flushbottom

\section{Introduction}

Neutrino oscillation experiments have provided the first evidence for new particle physics beyond the Standard Model (BSM) in the form of neutrino mass and mixing~\cite{nobel}. Although the origin of neutrino mass and mixing remains unknown~\cite{XZbook,King:2013eh}, there has been continuing experimental progress, for example on atmospheric mixing which is consistent with being maximal~\cite{Abe:2017uxa,Adamson:2017qqn,NOvA_new,Esteban:2016qun,deSalas:2017kay,Capozzi:2018ubv}.

The leading candidate for a theoretical explanation of neutrino mass and mixing remains the seesaw mechanism~\cite{Minkowski:1977sc,Yanagida:1979as,GellMann:1980vs,Schechter:1980gr,Mohapatra:1979ia,Mohapatra:1980yp}. However the seesaw mechanism involves a large number of free parameters at high energy, and is therefore difficult to test. One approach to reducing the number of seesaw parameters is to consider the minimal version involving only two right-handed neutrinos (2RHN)~\cite{King:1999mb, King:2002nf}. In such a scheme the lightest neutrino is massless. An early simplification~\cite{Frampton:2002qc} involved two texture zeros in the Dirac neutrino mass matrix consistent with cosmological leptogenesis~\cite{Fukugita:1986hr,Guo:2003cc, Ibarra:2003up, Mei:2003gn, Guo:2006qa, Antusch:2011nz,Harigaya:2012bw, Zhang:2015tea}. Although the normal hierarchy of neutrino masses, favoured by current data, is incompatible with the 2RHN model with two texture zeros~\cite{Harigaya:2012bw, Zhang:2015tea}, the one texture zero case originally proposed~\cite{King:1999mb, King:2002nf} remains viable. 

The Littlest Seesaw (LS) model is based on the 2RHN model with one texture zero, and in addition involves a well defined and constrained Yukawa involving just two independent Yukawa couplings~\cite{King:2013iva, Bjorkeroth:2014vha, King:2015dvf,Bjorkeroth:2015ora,Bjorkeroth:2015tsa,King:2016yvg,Ballett:2016yod}, leading to a highly predictive scheme. The LS model also provides a rather minimal explanation of cosmological leptogenesis, providing the lightest right-handed neutrino has a mass of order $10^{10}-10^{11}$ GeV~\cite{King:2018fqh}. However, to date there has been no attempt in the literature to address the origin of Dark Matter (DM) in the LS model.

The existence of DM in the Universe provides cosmological evidence for new physics beyond the Standard Model. Although there are many possible candidates for DM particles, it is interesting to try to connect the new physics required for DM particles to that required for neutrino mass and mixing, and there are many works in this direction~\cite{Caldwell:1993kn,Mohapatra:2002ug,Krauss:2002px,Ma:2006km,Asaka:2005an,Boehm:2006mi,Kubo:2006yx,Ma:2006fn,Hambye:2006zn,Lattanzi:2007ux,Ma:2007gq,Allahverdi:2007wt,Gu:2007ug,Sahu:2008aw,Arina:2008bb,Aoki:2008av,Ma:2008cu,Gu:2008yj,Aoki:2009vf,Gu:2010yf,Hirsch:2010ru,Esteves:2010sh,Kanemura:2011vm,Lindner:2011it,JosseMichaux:2011ba,Schmidt:2012yg,Borah:2012qr,Farzan:2012sa,Chao:2012mx,Gustafsson:2012vj,Blennow:2013pya,Law:2013saa,Hernandez:2013dta,Restrepo:2013aga,Chakraborty:2013gea,Ahriche:2014cda,Kanemura:2014rpa,Huang:2014bva,Varzielas:2015joa,Sanchez-Vega:2015qva,Fraser:2015mhb,Adhikari:2015woo,Ahriche:2016rgf,Sierra:2016qfa,Lu:2016ucn,Batell:2016zod,Ho:2016aye,Escudero:2016ksa,Bonilla:2016diq,Borah:2016zbd,Biswas:2016yan,Hierro:2016nwm,Bhattacharya:2016qsg,Chakraborty:2017dfg,Bhattacharya:2017sml,Ho:2017fte,Ghosh:2017fmr,Nanda:2017bmi,Narendra:2017uxl,Bernal:2017xat,Borah:2018gjk,Batell:2017cmf,Pospelov:2007mp,Falkowski:2009yz,Falkowski:2011xh,Cherry:2014xra,Bertoni:2014mva,Allahverdi:2016fvl,Karam:2015jta}. For example, a recent work considers the type-I seesaw mechanism together with a fermion singlet dark matter particle, stabilised by a discrete $Z_2$ symmetry arising from a broken $Z_4$~\cite{Bhattacharya:2018ljs}. The construction in Ref.~\cite{Bhattacharya:2018ljs} suggests that there exists a scalar field mediator between the two sectors whose vacuum expectation value not only generates the mass of the dark matter, but also takes part in the neutrino mass generation. This example is representative of many such attempts to connect the seesaw mechanism to DM, namely by invoking a Higgs portal type coupling to constrain the seesaw scale.

In this paper we propose a minimal and realistic model to simultaneously account for the neutrino masses through a type-I seesaw and provide a viable dark matter candidate which exists in a dark sector consisting of a single dark fermion $\chi$ and a single dark complex scalar $\phi$, both having an odd dark parity $Z_2$. Unlike many works in the literature, we focus on the right-handed neutrino portal (rather than the Higgs portal~\cite{Pospelov:2007mp,Burgess:2000yq,Davoudiasl:2004be,Bird:2006jd,Kim:2006af,Finkbeiner:2007kk,DEramo:2007anh,Barger:2007im,SungCheon:2007nw,MarchRussell:2008yu,McDonald:2008up,Piazza:2010ye,Pospelov:2011yp,Batell:2012mj,Kouvaris:2014uoa,Kainulainen:2015sva,Krnjaic:2015mbs,Tenkanen:2016jic,Heikinheimo:2016yds,Heikinheimo:2017ofk}) and on the implications of this for neutrino physics and dark matter phenomenology. In other words we suppose that the production of dark sector particles is achieved dominantly via their couplings to the right-handed neutrinos $N_R$, which are in turn coupled to the thermal bath via their neutrino Yukawa couplings to left-handed neutrinos and Higgs scalar. Thus the production of DM particles depends crucially on the same neutrino Yukawa couplings which determine neutrino mass and mixing. Another unique feature of our approach here is that we constrain the parameter space by the requirement of achieving a realistic pattern of neutrino masses and mixing, while many works in the literature only consider a toy model with a single neutrino mass for example. In order to achieve this, we focus on the LS model above with two right-handed neutrinos $N_R$, which successfully describes the neutrino data with a very small number of parameters. 

The requirement that the right-handed neutrino portal plays a dominant role in the production mechanism of dark sector particles, then provides a highly constrained and direct link between dark matter and neutrino mass and mixing. In order to achieve the correct relic density in this scenario we need to assume that the coupling of $N_R$ to the dark sector is very small, which puts us in the so called ``freeze-in'' scenario~\cite{Hall:2009bx} which is not usually considered in the framework of the right-handed neutrino portal (see Ref.~\cite{Bernal:2017kxu} for a recent detailed analysis on the freeze-in production mechanism). However in such a scenario, it becomes possible to obtain a successful relic density for almost arbitrarily large right-handed neutrino and dark sector masses. In particular, it is possible to achieve right-handed neutrino masses in the range  $10^{10}-10^{11}$ GeV required for leptogenesis. This means that the LS model can simultaneously achieve both leptogenesis and (when extended by a dark sector as here) a successful DM relic density. In such a case, the DM particles are ultra-heavy frozen-in particles, which we refer to as ``fimpzillas''.

The layout of the remainder of the paper is as follows. In Section~2 we discuss the Lagrangian of the model describing in detail the features of the LS model. In Section~3 we report the Boltzmann equations required to determine the DM relic abundance according to the freeze-in production mechanism. In Section~4 we discuss our numerical results, highlighting the regions of the parameter space that provide a link between DM and neutrino mass and mixing. Finally, in Section~5 we draw our conclusions.

\section {The model}

We consider an extension of the Littlest Seesaw model where the two right-handed neutrinos $N_{Ri}$ are coupled to a dark sector consisting of a scalar $\phi$ and a fermion $\chi$. For the sake of simplicity, we assume that the former is a complex field while the latter has a vector-like mass. The lighter particle between $\phi$ and $\chi$ plays the role of dark matter. The Standard Model (SM) and the Dark Sector (DS) are distinguished by a $Z_2$ symmetry that also stabilises the DM particles. In particular, the DS particles, $\phi$ and $\chi$, are odd under $Z_2$, while the other fields are even. In Tab.~\ref{tab:matter} we report the matter content of the model providing how the new fields transform under the electroweak gauge group $SU(2)_L \otimes U(1)_Y$ and the discrete symmetry $Z_2$.
\begin{table}[t!]
\centering
\begin{tabular}{|c|c|c|c|}
\hline
& $N_R$ & $\phi$ & $\chi$ \\ \hline \hline 
$SU(2)_L$ & {\bf 1} & {\bf 1} & {\bf 1} \\ \hline
$U(1)_Y$ & 0 & 0 & 0 \\ \hline  \hline
$Z_2$ & + & - & - \\ \hline
\end{tabular}
\caption{\label{tab:matter}New fields representations of the model, where $N_R$ are the two right-handed neutrinos, while
$\phi$ and $\chi$ are a new DS complex scalar and fermion, respectively.}
\end{table}
Hence, the full Lagrangian of the model can be divided in four parts
\begin{equation}
\mathcal{L} = \mathcal{L}_{\rm SM} + \mathcal{L}_{\rm Seesaw} + \mathcal{L}_{\rm DS} + \mathcal{L}_{\rm portal}\,,
\label{eq:lag}
\end{equation}
where the first term is the SM Lagrangian, the Seesaw term is responsible for neutrino masses, the DS part contains all the kinetic and mass terms of the dark particles $\phi$ and $\chi$, while the last term consists of the interactions that connect the visible and the dark sectors. In particular, the last three terms read
\begin{eqnarray}
\mathcal{L}_{\rm Seesaw} & = & - Y_{\alpha\beta} \overline{L_L}_\alpha \tilde{H} N_{R\beta} - \frac12 M_{R}\overline{N^c_{R}} N_{R} +h.c.\,, \label{eq:lagNS} \\
\mathcal{L}_{\rm DS} & = & \overline{\chi}\left(i \slashed{\partial} - m_\chi \right)\chi + \left|\partial_\mu \phi\right| - m^2_\phi \left|\phi\right|^2 + V\left(\phi\right)\,, \label{eq:lagDS} \\
\mathcal{L}_{\rm portal} & = & y_{\rm DS} \phi \, \overline{\chi}N_{R} + h.c \,, \label{eq:lagPortal}
\end{eqnarray}
where $L_{L\alpha}$ are the left-handed lepton doublets ($\alpha=e,\mu\tau$ and $\beta=1,2$) and
\begin{equation}
H = \left(\begin{array}{c}G^+ \\ \frac{v_{\rm SM} + h^0 + i G^0}{\sqrt2}\end{array}\right)
\end{equation}
is the SM Higgs doublet with $\tilde{H} = i \tau_2 H^*$. In Eq.~\eqref{eq:lagDS}, the quantities $m_\chi$ and $m_\phi$ are the masses of the fermion and the scalar, respectively, and $V\left(\phi\right)$ is a general potential for the scalar field allowed by the $Z_2$ symmetry. We assume that the discrete $Z_2$ symmetry is an exact symmetry of the model so that the scalar field does not acquire a v.e.v. (a detailed analysis of the scalar potential is beyond the scope of this paper). Finally, the visible and dark sectors are connected through the right-handed neutrino portal defined in Eq.~\eqref{eq:lagPortal}, where for the sake of simplicity we have assumed the same real coupling $y_{\rm DS}$ between the two right-handed neutrinos and the DS particles. It is worth noticing that in this framework the Higgs portal defined by the coupling $y_{H\phi} \left|H\right|^2\left|\phi\right|^2$ is also allowed. However, the aim of the present analysis is to investigate the impact of the right-handed neutrino portal on the DM phenomenology and to highlight the interesting connection between neutrinos and DM particles. Hence, for this reason we consider the coupling $y_{H\phi}$ to be negligible. 

Let us now discuss in detail the Littlest Seesaw model defined by the Lagrangian given in Eq.~\eqref{eq:lagNS}. The first term is a Yukawa-like coupling while the second one is the Majorana mass term for the right-handed neutrinos. After the electroweak symmetry breaking, the light effective left-handed Majorana neutrino mass matrix is obtained by type-I seesaw formula~\cite{Minkowski:1977sc,Yanagida:1979as,GellMann:1980vs,Schechter:1980gr,Mohapatra:1979ia,Mohapatra:1980yp}
\begin{equation}
m_\nu = -m_D M^{-1}_R {m_D}^T \,,
\label{eq:ssformula}
\end{equation}
where the neutrino Dirac mass matrix $m_D$ is defined as
\begin{equation}
m_D = \frac{v_{\rm SM}}{\sqrt 2} Y\,,
\label{mD1}
\end{equation}
being $v_{\rm SM}=246$~GeV the SM Higgs v.e.v. In the LS model~\cite{King:2013iva, Bjorkeroth:2014vha, King:2015dvf,Bjorkeroth:2015ora,Bjorkeroth:2015tsa,King:2016yvg,Ballett:2016yod}, in the basis where the right-handed neutrino Majorana mass matrix is diagonal,
\begin{equation}
M_R =  \left(\begin{array}{cc} M_{\rm R1} &  0 \\ 0 & M_{\rm R2} \end{array}\right)\,,
\end{equation}
the neutrino Dirac mass matrix has the form in the LR convention
\begin{equation}
m_D = \left(\begin{array}{lr} 0 &  b \, e^{i \frac{\eta}{2}} \\ a & 3b \, e^{i \frac{\eta}{2}} \\ a & b \, e^{i \frac{\eta}{2}} \end{array}\right) \,,
\label{mD2}
\end{equation}
while the neutrino Majorana mass matrix for the light neutrinos $\nu_L$ is given effectively in the seesaw approximation~\eqref{eq:ssformula} by
\begin{equation}
m_\nu = m_a \left(\begin{array}{ccc} 0 & 0 & 0 \\ 0 & 1 & 1 \\ 0 & 1 & 1 \end{array}\right) + m_b \, e^{i \eta} \left(\begin{array}{ccc} 1 & 3 & 1 \\ 3 & 9 & 3 \\ 1 & 3 & 1 \end{array}\right)\,.
\label{eq:numass}
\end{equation}
In the above expressions, $a$ and $b$ are two real couplings while $\eta$ is their relative phase, and
\begin{equation}
m_a = \frac{a^2}{M_{R1}} \qquad {\rm and} \qquad m_b = \frac{b^2}{M_{R2}} \,.
\label{mD3}
\end{equation}
Hence, in the LS model the Yukawa matrix in Eq.~\eqref{eq:lagNS} reads
\begin{equation}
Y = \sqrt{\frac{2 \,m_a\, M_{\rm R1}}{v^2_{SM}}} \left(\begin{array}{cc} 0 & 0 \\ 1 & 0 \\ 1 & 0\end{array}\right) + \sqrt{\frac{2 \,m_b\, M_{\rm R2}}{v^2_{SM}}} e^{i \frac{\eta}{2}} \left(\begin{array}{cc} 0 & 1 \\ 0 & 3  \\ 0 & 1\end{array}\right)\,.
\label{eq:yuk}
\end{equation}
The result in Eq.~\eqref{eq:yuk} follows from Eqs.~\eqref{mD1},~\eqref{mD2}~and~\eqref{mD3}. From a model building perspective, it may be achieved by the effective  Yukawa coupling of the first right-handed neutrino having an alignment proportional to $\left(0,1,1\right)$, while that of the second right-handed neutrino having an alignment proportional to $\left(1,3,1\right)$. The theoretical origin of such alignments may be related to a spontaneously broken family symmetry under which the three families of electroweak lepton doublets transform as an irreducible triplet, as discussed further in Refs.~\cite{King:2013iva, Bjorkeroth:2014vha, King:2015dvf,Bjorkeroth:2015ora,Bjorkeroth:2015tsa,King:2016yvg,Ballett:2016yod}. The phenomenological desirability of such a structure, from the point of view of both low energy neutrino data and leptogenesis, was recently discussed in Ref.~\cite{King:2018fqh}.

This minimal framework allows for a very good fit of neutrino mixing and mass parameters. In particular, according to Ref.~\cite{Ballett:2016yod} we consider the following benchmark values for the three parameters that provide a nice agreement with the experimental neutrino measurements:
\begin{equation}
m_a = 26.74~{\rm \,meV}\,, \qquad m_b = 2.682~{\rm \,meV}\,, \qquad {\rm and} \qquad \eta = \frac23\pi\,.
\label{eq:nu-data}
\end{equation}
Hence in the model, there are five free parameters: the two right-handed neutrino masses, the two masses $m_\chi$ and $m_\phi$, and the right-handed neutrino portal coupling $y_{\rm DS}$. However, for the sake of simplicity, we consider the case where the two right-handed neutrinos have the same mass $M_{R1} = M_{R2} = M_{R}$, where $M_{R}$ is then identified with the energy scale of the seesaw. The other quantities are then fixed by the experimental neutrino measurements according to Eq.~\eqref{eq:nu-data}.  In the following, we focus on the case $m_\chi \leq m_\phi$ implying that the fermions $\chi$ are stable and play the role of DM particles. The other case is a trivial modification and, therefore, is not discussed here. In the next Section, we report the Boltzmann equations important for the DM production in the early Universe.

\section{The Boltzmann equations in freeze-in dark matter production}

The Boltzmann equations encode how the yield $Y_i$ of particles of species $i$ evolves with the temperature $T$. The most general expression takes the form~\cite{Kolb:1990vq}\footnote{The Botlzmann equations are here written in terms of the temperature $T$ rather than the time or the auxiliary variable $x = M_R/T$. This implies different signs in the right-hand side of the Boltzmann equations because the photon temperature decreases during the evolution of the Universe.}
\begin{eqnarray}
\mathcal{H}\,T\left(1+\frac{T}{3 g^\mathfrak{s}_*\left(T\right)}\frac{d g^\mathfrak{s}_*}{d T}\right)^{-1}\frac{d Y_i}{d T} &=& \sum_{kl} \left<\Gamma_{i\rightarrow kl}\right>Y_i^{\rm eq}\left(\frac{Y_i}{Y_i^{\rm eq}}-\frac{Y_k \, Y_l}{Y_k^{\rm eq}Y_l^{\rm eq}}\right)	\nonumber \\
&& - \sum_{jk} \left<\Gamma_{j\rightarrow ik}\right>Y_j^{\rm eq}\left(\frac{Y_j}{Y_j^{\rm eq}}-\frac{Y_i \, Y_k}{Y_i^{\rm eq}Y_k^{\rm eq}}\right) \\
&&+ \mathfrak{s} \sum_{jkl} \left<\sigma_{ij\rightarrow kl}\, v_{ij}\right> Y_i^{\rm eq}Y_j^{\rm eq}\left(\frac{Y_i \, Y_j}{Y_i^{\rm eq}Y_j^{\rm eq}}-\frac{Y_k \, Y_l}{Y_k^{\rm eq}Y_l^{\rm eq}}\right) \,, \nonumber
\label{eq:Boltz}
\end{eqnarray}
where $\mathcal{H}$ and $\mathfrak{s}$ are, respectively, the Hubble parameter and the entropy density of the thermal bath\footnote{We use the symbol $ \mathfrak{s}$ to represent entropy density and reserve the symbol $s$ for the Mandelstam variable.}
\begin{equation}
\mathcal{H} = 1.66 \sqrt{g_*\left(T\right)}\frac{T^2}{M_{\rm Planck}} \qquad{\rm and}\qquad \mathfrak{s} = \frac{2\pi^2}{45} g^\mathfrak{s}_*\left(T\right) T^3 \,,
\label{eq:hubble}
\end{equation}
where $M_{\rm Planck}=1.22\times10^{19}$~GeV is the Planck mass, and $g_*$ and $g^\mathfrak{s}_*$ are the degrees of freedom of the relativistic species in the thermal bath. In the Boltzmann equation~\eqref{eq:Boltz}, the quantity $Y^{\rm eq}_i$ is the yield at the thermal equilibrium that takes the expression
\begin{equation}
Y^{\rm eq}_{i} \equiv \frac{n^{\rm eq}_{i}}{\mathfrak{s}}\qquad{\rm with}\qquad n^{\rm eq}_{i} = \frac{g_i \, m_i^2 \, T}{2\pi^2} K_2\left(\frac{m_i}{T}\right)\,,
\label{eq:yeq}
\end{equation}
where $m_i$ and $g_i$ are respectively the mass and the internal degrees of freedom of particles $i$, and $K_2$ denotes the modified Bessel function of the second kind of order 2. The Boltzmann equation~\eqref{eq:Boltz} takes into account all the decay processes $i\rightarrow kl$ and $j\rightarrow ik$ (first and second terms) and all the scattering ones $ij\rightarrow kl$ (third term). In particular, the thermally averaged decay width is given by
\begin{equation}
\left< \Gamma_{i\rightarrow kl} \right> = \frac{K_1\left(m_i/T\right)}{K_2\left(m_i/T\right)}\Gamma_{i\rightarrow kl}\,,
\end{equation}
with $K_1$ being the modified Bessel function of the second kind of order 1, while the thermally averaged cross section takes the form~\cite{Edsjo:1997bg}
\begin{equation}
\left<\sigma_{ij\rightarrow kl}\, v_{ij}\right> = \frac{1}{n^{\rm eq}_{i}\, n^{\rm eq}_{j}}\frac{g_i \, g_j}{S_{kl}}\frac{T}{512\pi^6} \int_{\left(m_i+m_j\right)^2}^\infty ds \, \frac{p_{ij} \, p_{kl} \,K_1\left(\sqrt{s}/T\right)}{\sqrt{s}} \int  \overline{\left|\mathcal{M}\right|^2}_{ij\rightarrow kl} \, d\Omega\,.
\end{equation}
where $s$ is the Mandelstam variable equal to the 
square of the centre-of-mass energy, $S_{kl}$ is a symmetry factor, $p_{ij}$ ($p_{kl}$) is the initial (final) centre-of-mass momentum and $\overline{\left|\mathcal{M}\right|^2}$ is the averaged squared amplitude evaluated in the centre-of-mass frame.

In order to compute the DM relic abundance, we need in principle to solve four coupled Boltzmann equations: two equations for the two right-handed neutrinos $n_1$ and $n_2$, one for the dark scalars $\phi$ and one for the DM particles $\chi$. However, in all the cases analysed, we consider the masses of the four new particles larger than the electroweak scale. In this limit, the neutrino Yukawa couplings given in Eq.~\eqref{eq:yuk} are large enough to bring the two right-handed neutrino in thermal equilibrium with the thermal bath. Hence, their yield follows the thermal distribution defined in Eq.~\eqref{eq:yeq}. This allows us to solve just the two coupled Boltzmann equations for the particles in the dark sector. Furthermore, we also have $g^\mathfrak{s}_* = g^\mathfrak{s}_{*n} + g^\mathfrak{s}_{*\rm SM}$ for $T > M_R$ and $g^\mathfrak{s}_* = g^\mathfrak{s}_{*\rm SM}$ for $T < M_R$, with $g^\mathfrak{s}_{*n} = 2\times2\times7/8$ and $g^\mathfrak{s}_{*\rm SM} = 106.75$ that corresponds to the total number of relativistic degrees of freedom in the SM at temperature above the electroweak scale. In both limits we have $g_* = g^\mathfrak{s}_*$.

In the present analysis, we consider the right-handed neutrino portal coupling $y_{\rm DS}$ to be very small. Only in this case, the neutrino sector defined by the Yukawa couplings in Eq.~\eqref{eq:yuk} could play an important role in DM production. Indeed, due to the smallness of the neutrino Yukawa couplings fixed by the seesaw formula~\eqref{eq:ssformula}, the scatterings involving active neutrinos and leptons could dominate over the other processes in the DM production only when the right-handed neutrino portal is suppressed by a very small coupling as well. When this occurs, we would have an interesting relation between neutrino physics and dark matter. In the limit $y_{\rm DS} \ll 1$, DM particles are produced through the so-called {\it freeze-in} production mechanism~\cite{Hall:2009bx}. Differently from the standard {\it freeze-out} production mechanism~\cite{Lee:1977ua,Bernstein:1985th,Gondolo:1990dk}, the DM particles are never in thermal equilibrium and are instead gradually produced from the thermal bath through very weak interactions, while the rate of the back-reactions that would destroy DM particles are further suppressed since $Y_{\chi,\phi} \ll Y_{\chi,\phi}^{\rm eq}$ at high temperatures. On the other hand, in the freeze-out regime the coupling $y_{\rm DS}$ has to be assumed large enough to bring the dark sector in thermal equilibrium, spoiling the neutrino-dark matter relation.

Moreover, in the freeze-in scenario, since the dark sector is almost decoupled with the SM one ($y_{\rm DS} \ll 1$), all the constraints coming from collider, direct and indirect DM searches are practically circumvented. As will be discussed later, in the range of masses considered the correct DM abundance is achieved for a right-handed neutrino portal coupling smaller than $10^{-4}$. Consequently, since the DM annihilation processes important for indirect DM searches are in general suppressed by $y_{\rm DS}^4$, the corresponding indirect signals are very small. On the other hand, due to the absence of a direct coupling to quarks, the processes important for collider and direct DM searches are even more suppressed.

Once the two Boltzmann equations are solved, the total DM relic abundance is then given by
\begin{equation}
\Omega_{\rm DM}h^2 = \frac{\rho_{\rm DM,0}}{\rho_{\rm crit}/h^2}\,,
\label{eq:omegaPRE}
\end{equation}
where $\rho_{\rm DM,0}$ is today’s energy density of DM particles and $\rho_{\rm crit}/h^2 = 1.054 \times10^{-5}~{\rm GeV \, cm^{-3}}$ is the critical density~\cite{Patrignani:2016xqp}. In oder to provide a realistic model for viable DM candidates, this quantity has to be equal to its experimental value that has been measured by Planck Collaboration at 68\%~C.L.~\cite{Ade:2015xua}:
\begin{equation}
\left.\Omega_{\rm DM}h^2\right|_{\rm obs} = 0.1188 \pm 0.0010\,.
\label{eq:omegaOBS}
\end{equation}

In this framework, by considering the simple case of degenerate right-handed neutrinos ($M_{R1} = M_{R2} = M_R$) one may classify two different types of orderings among the masses of new particles:
\begin{itemize}
\item {\bf Ordering type A:} $m_\phi \geq M_R,\, m_\chi$;
\item {\bf Ordering type B:} $M_R \geq m_\phi,\, m_\chi$.
\end{itemize}
 In all the cases, the fermion particles $\chi$ are the lightest dark particles and play the role of dark matter. As it will be clear later, the main difference in the DM production for the two hierarchies resides in the fact that the two-body decay of right-handed neutrinos into dark particles is allowed or not depending on the mass ordering. Therefore the results are insensitive to the detailed sub-orderings within the above classification. Before we discuss in detail the expressions of the two Boltzmann equations for the two different types of ordering, it is worth observing that other more involved orderings are allowed if one relaxes the assumption of having degenerate right-handed neutrinos. However, all the following considerations can be applied for just one right-handed neutrino at a time, according to its mass ordering with respect to the masses of dark particles. For this reason, we prefer to focus on the simplest case of equal right-handed neutrino masses.

\subsection{Ordering type A: $m_\phi \geq M_R,\, m_\chi$}

In this case, the Boltzmann equations for the two dark species are given by
\begin{eqnarray}
\mathcal{H}\,T\left(1+\frac{T}{3 g^\mathfrak{s}_*\left(T\right)}\frac{d g^\mathfrak{s}_*}{d T}\right)^{-1} \frac{d Y_\phi}{d T} & = & - \mathfrak{s} \left<\sigma\, v\right>_{\phi\phi}^{\rm DS} \left({Y_\phi^{\rm eq}}\right)^2 - \mathfrak{s} \left<\sigma\, v\right>^{\rm \nu-Yukawa}_{\chi\phi} Y_\phi^{\rm eq}Y_\chi^{\rm eq} \nonumber \\
& & + \left<\Gamma_{\phi}\right>\left(Y_\phi-\frac{Y_\phi^{\rm eq}}{Y_\chi^{\rm eq}}Y_\chi\right)  \,, \label{eq:phi1} \\
\mathcal{H}\,T\left(1+\frac{T}{3 g^\mathfrak{s}_*\left(T\right)}\frac{d g^\mathfrak{s}_*}{d T}\right)^{-1} \frac{d Y_\chi}{d T} & = & - \mathfrak{s}  \left<\sigma\, v\right>_{\chi\chi}^{\rm DS} \left({Y_\chi^{\rm eq}}\right)^2 - \mathfrak{s} \left<\sigma\, v\right>^{\rm \nu-Yukawa}_{\chi\phi} Y_\phi^{\rm eq}Y_\chi^{\rm eq}  \nonumber \\
& & -  \left<\Gamma_{\phi}\right>\left(Y_\phi-\frac{Y_\phi^{\rm eq}}{Y_\chi^{\rm eq}}Y_\chi\right) \,, \label{eq:chi1}
\end{eqnarray}
where the right-handed neutrinos $N_R$ are taken to be in thermal equilibrium with photons. In the above expressions, we have neglected all the other subdominant terms suppressed by the condition $Y_{\chi,\phi} \ll Y_{\chi,\phi}^{\rm eq}$ according to the freeze-in production paradigm. Moreover,
we do not take into account all the other scattering processes with right-handed neutrinos that are suppressed by the active-sterile neutrino mixing $\theta\equiv m_D M^{-1}_R$ (see Refs.~\cite{Buchmuller:1990vh,Pilaftsis:1991ug}). This is indeed a good approximation in the case where the right-handed neutrinos are heavier than the electroweak energy scale. Hence, we can mainly distinguish three different classes of processes (see Fig.~\ref{fig:Feyn1}), whose expressions are reported in the Appendix.
\begin{figure}[t!]
\begin{center}
\subfigure[Dark Sector scatterings]{\begin{fmffile}{DS1}
\fmfframe(15,15)(15,15){
\begin{fmfgraph*}(85,40)
\fmflabel{$\phi^*$}{i1}
\fmflabel{$\phi$}{i2}
\fmflabel{$n_j$}{o1}
\fmflabel{$n_i$}{o2}
\fmfv{label=$y_{\rm DS}$}{v1}
\fmfv{label=$y_{\rm DS}$}{v2}
\fmfleft{i1,i2}
\fmfright{o1,o2}
\fmf{dashes}{i2,v1}
\fmf{plain}{v1,o2}
\fmf{dashes}{i1,v2}
\fmf{plain}{v2,o1}
\fmf{plain,label=$\chi$,tension=0}{v1,v2}
\fmfdotn{v}{2}
\end{fmfgraph*}}
\end{fmffile}
\begin{fmffile}{DS2}
\fmfframe(15,15)(15,15){
\begin{fmfgraph*}(85,40)
\fmflabel{$\overline{\chi}$}{i1}
\fmflabel{$\chi$}{i2}
\fmflabel{$n_j$}{o1}
\fmflabel{$n_i$}{o2}
\fmfv{label=$y_{\rm DS}$}{v1}
\fmfv{label=$y_{\rm DS}$}{v2}
\fmfleft{i1,i2}
\fmfright{o1,o2}
\fmf{plain}{i2,v1}
\fmf{plain}{v1,o2}
\fmf{plain}{i1,v2}
\fmf{plain}{v2,o1}
\fmf{dashes,label=$\phi$,tension=0}{v1,v2}
\fmfdotn{v}{2}
\end{fmfgraph*}}
\end{fmffile}}
\subfigure[Neutrino Yukawa scatterings]{\begin{fmffile}{NS1}
\fmfframe(20,15)(15,15){
\begin{fmfgraph*}(85,40)
\fmflabel{$\phi,\phi^*$}{i1}
\fmflabel{$\overline{\chi},\chi$}{i2}
\fmflabel{$\nu_i,\nu_i,\ell_i^\pm$}{o2}
\fmflabel{$h^0,G^0,G^\mp$}{o1}
\fmfv{label=$y_{\rm DS}$}{v1}
\fmfv{label=$y_\nu$}{v2}
\fmfleft{i1,i2}
\fmfright{o1,o2}
\fmf{plain}{i2,v1}
\fmf{dashes}{i1,v1}
\fmf{dashes}{v2,o1}
\fmf{plain}{v2,o2}
\fmf{plain,label=$n_j$}{v1,v2}
\fmfdotn{v}{2}
\end{fmfgraph*}}
\end{fmffile}}
\end{center}
\caption{\label{fig:Feyn1}Dominant scattering processes responsible for DM production in the case of ordering type A. Here $n_i$ represent the heavy neutrino mass eigenstates (dominantly from the right-handed neutrinos $N_R$) while $\nu_i$ represent the light neutrino mass eigenstates (dominantly from $\nu_L$ in the doublet $L_L$). }
\end{figure}
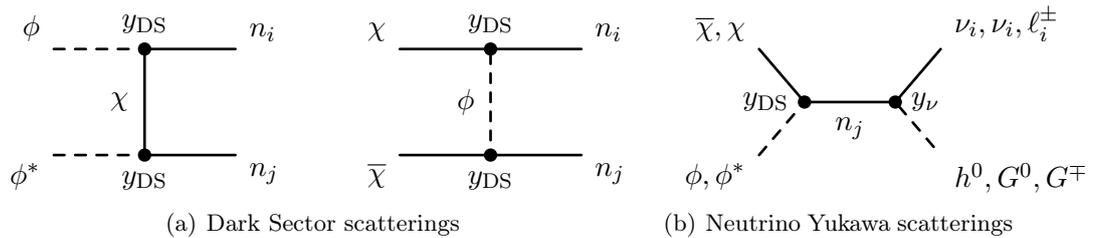
In particular, we have:
\begin{itemize}
\item {\bf Dark Sector scatterings:} the first term in the two Boltzmann equations refers to the scattering processes $\phi\phi^*\rightarrow n_in_j$ and $\chi\overline{\chi}\rightarrow n_in_j$, respectively. In particular, we have
\begin{eqnarray}
\left<\sigma\, v\right>_{\phi\phi}^{\rm DS} & = & \sum_{i,j=1,2} \left<\sigma_{\phi\phi^*\rightarrow n_in_j}\, v\right> \,, \\
\left<\sigma\, v\right>_{\chi\chi}^{\rm DS} & = & \sum_{i,j=1,2} \left<\sigma_{\chi\overline{\chi}\rightarrow n_in_j}\, v\right> \,.
\end{eqnarray}
The amplitudes of such processes are obtained with different contractions of the right-handed neutrino portal in Eq.~\eqref{eq:lagPortal} and, therefore, the corresponding thermal averaged cross sections are proportional to the coupling $y_{\rm DS}^4$;
\item {\bf Neutrino Yukawa scatterings:} the second term in both equations corresponds to the scattering processes that originate from the neutrino Yukawa interaction given in Eq.~\eqref{eq:lagNS}. According to the Goldstone Boson Equivalence Theorem, we consider also the processes involving the other degrees of freedom of the Higgs doublet, $G^0$ and $G^\pm$. Hence, we have
\begin{equation}
\left<\sigma\, v\right>^{\rm \nu-Yukawa}_{\chi\phi} = \sum_{i=1}^3 \left[\left<\sigma_{\chi \phi\rightarrow \nu_i h^0}\, v\right> + \left<\sigma_{\chi \phi\rightarrow \nu_i G^0}\, v\right> + \left<\sigma_{\chi \phi\rightarrow \ell^\pm_i G^\mp}\, v\right>\right]\,.
\end{equation}
The size of such $\nu-$Yukawa processes is set by the product $y_{\rm DS}^2\left|y_\nu\right|^2$, where $y_\nu$ is proportional to the Yukawa couplings in Eq.~\eqref{eq:yuk}. In the neutrino mass basis, we have $y_\nu = \left(U^\dagger_\nu Y\right)_{ij}/\sqrt2$ or $y_\nu = Y_{ij}$ for neutrinos or leptons in the final states, respectively. The matrix $U_\nu$ is the Pontecorvo–Maki–Nakagawa–Sakata matrix describing neutrino oscillations. In the present analysis, we consider the best-fit values of neutrino oscillation parameters reported by the latest neutrino global analysis~\cite{Esteban:2016qun} (see also Refs.~\cite{Capozzi:2018ubv,deSalas:2017kay});
\item {\bf Scalar decay: } the last terms are related to the two-body decays of $\phi$ particles into $\chi$ and right-handed neutrino, which are kinematically allowed if $m_\phi > m_\chi + M_R$. The corresponding partial decay widths are proportional to $y_{\rm DS}^2$ and take the expression
\begin{equation}
\Gamma_{\phi \rightarrow \chi n_i}^{\rm 2-body} = \frac{y_{\rm DS}^2\,m_\phi}{16 \pi}\left(1-\frac{m_\chi^2}{m_\phi^2}-\frac{M_R^2}{m_\phi^2}\right)\lambda\left(1,\frac{m_\chi}{m_\phi},\frac{M_R}{m_\phi}\right) \,,
\end{equation}
where $\lambda$ is the Kallen function. The total decay width is simply given by the sum of the two widths, i.e.
\begin{equation}
\Gamma_{\phi}^{\rm 2-body} = \sum_{i=1}^2 \Gamma_{\phi \rightarrow \chi n_i}^{\rm 2-body} \,.
\end{equation}
\end{itemize}
It is worth observing that the DS and $\nu-$Yukawa scattering processes scale in a different way with the right-handed neutrino portal coupling $y_{\rm DS}$. As will be discussed in detail, this implies that there are regions of the parameter space where one of the two different processes dominates in the DM production. While the two scattering processes are responsible for the production of particles in the dark sector, the decays allow to convert all the dark scalars $\phi$ into DM particles $\chi$. Indeed, if $m_\phi > m_\chi + M_R$, all the dark scalars particles are converted into the lighter fermions according to the discrete $Z_2$ symmetry. These scalar two-body decays are in general faster than the Universe expansion set by the Hubble parameter and occur at temperature higher than the electroweak energy scale. Hence, in case of $m_\phi > m_\chi + M_R$,  the total DM relic abundance can be cast in terms of an effective DM yield defined as
\begin{equation}
Y_{\rm DM} \left(T\right)= Y_\chi \left(T\right) + Y_\phi \left(T\right) \,,
\end{equation}
which then obeys to the following effective Boltzmann equation
\begin{eqnarray}
\mathcal{H}\,T\left(1+\frac{T}{3 g^\mathfrak{s}_*\left(T\right)}\frac{d g^\mathfrak{s}_*}{d T}\right)^{-1} \frac{d Y_{\rm DM}}{d T} & = & - \mathfrak{s} \left<\sigma\, v\right>_{\phi\phi}^{\rm DS} \left({Y_\phi^{\rm eq}}\right)^2 - \mathfrak{s} \left<\sigma\, v\right>_{\chi\chi}^{\rm DS}  \left({Y_\chi^{\rm eq}}\right)^2 \nonumber \\
& & - 2 \, \mathfrak{s} \left<\sigma\, v\right>^{\rm \nu-Yukawa}_{\chi\phi} Y_\phi^{\rm eq}Y_\chi^{\rm eq}\,.
\end{eqnarray}
This differential equation can be easily integrated providing the today's DM yield at $T=0$, denoted as $Y_{\rm DM,0}$. Such a quantity is given by the sum of three different contributions related to DS and $\nu-$Yukawa processes:
\begin{equation}
Y_{\rm DM,0} = Y^{\rm DS}_\phi + Y^{\rm DS}_\chi + 2\,Y^{\rm \nu-Yukawa}\,,
\label{eq:DMyield}
\end{equation}
where
\begin{eqnarray}
Y^{\rm DS}_\phi & = &  \int_0^\infty dT \, \frac{\mathfrak{s}}{\mathcal{H}\,T} \left(1+\frac{T}{3 g^\mathfrak{s}_*\left(T\right)}\frac{d g^\mathfrak{s}_*}{d T}\right) \left<\sigma\, v\right>_{\phi\phi}^{\rm DS} \left({Y_\phi^{\rm eq}} \right)^2 \,, \label{eq:1}\\
Y^{\rm DS}_\chi & = &  \int_0^\infty dT \, \frac{\mathfrak{s}}{\mathcal{H}\,T} \left(1+\frac{T}{3 g^\mathfrak{s}_*\left(T\right)}\frac{d g^\mathfrak{s}_*}{d T}\right)  \left<\sigma\, v\right>_{\chi\chi}^{\rm DS} \left({Y_\chi^{\rm eq}}\right)^2 \,, \label{eq:2}\\
Y^{\rm \nu-Yukawa} & = & \int_0^\infty dT \, \frac{\mathfrak{s}}{\mathcal{H}\,T} \left(1+\frac{T}{3 g^\mathfrak{s}_*\left(T\right)}\frac{d g^\mathfrak{s}_*}{d T}\right) \left<\sigma\, v\right>^{\rm \nu-Yukawa}_{\chi\phi}Y_\phi^{\rm eq}Y_\chi^{\rm eq} \,. \label{eq:3}
\end{eqnarray}
Here, we have assumed negligible yields of dark particles as initial condition. We note that the above expressions are dimensionless. However, we can rewrite Eq.~\eqref{eq:DMyield} in a way that one can explicitly see how it depends on the seesaw energy scale $M_R$, the right-handed neutrino portal coupling $y_{\rm DS}$ and the Yukawa couplings $y_\nu$ evaluated at the GeV energy scale. In particular, we have
\begin{equation}
Y_{\rm DM,0} = y_{\rm DS}^4\left(\frac{1~{\rm GeV}}{M_R}\right) \left[ \tilde{Y}^{\rm DS}_\phi + \tilde{Y}^{\rm DS}_\chi \right] + y_{\rm DS}^2 \tilde{y}_\nu^2\left(1~{\rm GeV}\right) \left[ 2 \, \tilde{Y}^{\rm \nu-Yukawa} \right] \,,
\label{eq:DMyieldDimLess}
\end{equation}
where the dimensionless quantities $\tilde{Y}$ are suitably defined by using Eqs.~\eqref{eq:1},~\eqref{eq:2} and~\eqref{eq:3}, and the effective squared Yukawa coupling $\tilde{y}_\nu^2$ is equal to (see the Appendix)
\begin{equation}
\tilde{y}_\nu^2 \left(M_R\right) = 2.47 \times 10^{-14} \left(\frac{M_R}{\rm 1~GeV}\right)\,.
\label{eq:effyuk}
\end{equation}
We note once again that, due to the different dependence on the coupling $y_{\rm DS}$ and the seesaw scale $M_R$ (see Eq.~\eqref{eq:yuk}), one of the two different classes of scattering processes could dominate over the other one. Hence, we expect that the neutrino sector defined by the Yukawa couplings in Eq.~\eqref{eq:yuk}, whose structure could be inferred by low-energy neutrino experiments, would drive the DM production in some regions of the parameter space. This implies an intriguing relation between neutrino physics and dark matter.

On the other hand, in case of $m_\phi \leq m_\chi + M_R$ the scalar decays are not kinematically allowed and both dark particles are stable, providing a {\it two-component} DM scenario. By integrating the two Boltzmann equations, we have that the two today's yields of $\chi$ and $\phi$ particles are given by
\begin{equation}
Y_{\chi,0} = Y^{\rm DS}_\chi + Y^{\rm \nu-Yukawa}\,,\qquad{\rm and}\qquad Y_{\phi,0} = Y^{\rm DS}_\phi + Y^{\rm \nu-Yukawa}\,.
\end{equation}
According to Eq.~\eqref{eq:omegaPRE}, the total DM relic abundance predicted by the model takes the form
\begin{equation}
\Omega_{\rm DM}h^2 =\left\{ \begin{array}{lcl} \frac{2 \, \mathfrak{s}_0 \, m_\chi  \,Y_{\rm DM,0}}{\rho_{\rm crit}/h^2} & \qquad & {\rm for} \,\,\,\, m_\phi > m_\chi + M_R \\
& \qquad &  \\ 
\frac{2 \, \mathfrak{s}_0 \, \left(m_\chi  \,Y_{\chi,0} + m_\phi  \,Y_{\phi,0} \right)}{\rho_{\rm crit}/h^2} & \qquad & {\rm for} \,\,\,\, m_\phi \leq m_\chi + M_R \end{array}\right.\,,
\label{eq:omegaPREA}
\end{equation}
where $\mathfrak{s}_0=2891.2\,{\rm cm^3}$ is today’s entropy density~\cite{Patrignani:2016xqp}, and the factor of 2 takes into account the contribution of DM anti-particles. Before concluding, we remind that all the above equations and expressions still hold in both the two possible hierarchies $M_R \leq m_\chi \leq m_\phi$ and $m_\chi \leq M_R \leq m_\phi$, once the different phase space provided by the different mass hierarchy is taken into account.

\subsection{Ordering type B: $M_R \geq m_\phi,\, m_\chi$}

In the ordering B case, the dark particles are mainly produced through the two-body decays of the two right-handed neutrinos $n_i \rightarrow \chi \phi^*,\phi\overline{\chi}$. The decay widths take the expression
\begin{equation}
\Gamma_{n_i\rightarrow \chi \phi^*} = \frac{y_{\rm DS}^2\,M_R}{32 \pi}\left(1+\frac{m_\chi^2}{M_R^2}-\frac{m_\phi^2}{M_R^2}\right)\lambda\left(1,\frac{m_\chi}{M_R},\frac{m_\phi}{M_R}\right)\,.
\end{equation}
Since the corresponding term in the Boltzmann equation is just proportional to $y_{\rm DS}^2$, such two-body decays provide the dominant contribution to the DM production. On the other hand, the contributions coming from the DS and $\nu-$Yukawa processes are almost sub-dominant. Moreover, in this scenario the dark scalar particles $\phi$ are slowly converted into the fermion ones $\chi$ through three-body decays with virtual right-handed neutrinos. In the limit $m_\phi \gg m_\chi$, the scalar three-body decay width is given by
\begin{equation}
\Gamma_\phi^{\rm 3-body} = \frac{y_{\rm DS}^2\tilde{y}_\nu^2}{1536 \pi^3}\frac{m_\phi^3}{M^2_R}\left(1+\frac{m_\phi^2}{2M_R^2}\right)\,,
\label{eq:dec3}
\end{equation}
where $\tilde{y}_\nu^2$ is the effective squared Yukawa coupling defined in Eq.~\eqref{eq:effyuk}. Such decays, which are suppressed by the small Yukawa coupling and by the right-handed neutrino mass, only occur at very late times. Hence, the ordering B case would lead to a two-component dark matter scenario where both dark scalars and fermions contribute to the final DM relic abundance. In this case, the two Boltzmann equations further simplifies and the today's yields of $\chi$ and $\phi$ are both equal to
\begin{equation}
Y_{\rm \chi,0} = Y_{\rm \phi,0} = \int_0^\infty dT \, \frac{1}{\mathcal{H}\,T} \sum_{i=1,2} \left[\left<\Gamma_{n_i\rightarrow \chi \phi^*} \right> Y^{\rm eq}_{n_i} \right] = \left(\frac{135 \, M_{\rm Planck}}{1.66 \left(2 \pi^3\right) g^\mathfrak{s}_* \sqrt{g_*}}\right)\frac{\Gamma_{n_i\rightarrow \chi \phi^*}}{M^2_R} \,,
\label{eq:DMC}
\end{equation}
and the final DM relic abundance is then given by
\begin{equation}
\Omega_{\rm DM}h^2 = \frac{2 \, \mathfrak{s}_0 \left(m_\chi Y_{\rm \chi,0} + m_\phi Y_{\rm \phi,0} \right) }{\rho_{\rm crit}/h^2} \,.
\end{equation}
By comparing the above expression with Eq.~\eqref{eq:omegaOBS} and using Eq.~\eqref{eq:DMC}, we obtain the following analytical expression for the right-handed neutrino portal coupling $y_{\rm DS}$ required to account for the correct DM relic abundance,
\begin{equation}
y_{\rm DS} = 1.22\times 10^{-12} \left(\frac{g^\mathfrak{s}_*}{106.75}\right)^{3/2} \sqrt{\frac{M_R}{m_\phi+m_\chi}} \left[\left(1+\frac{m_\chi^2}{M_R^2}-\frac{m_\phi^2}{M_R^2}\right)\lambda\left(1,\frac{m_\chi}{M_R},\frac{m_\phi}{M_R}\right)\right]^{-\frac12}\,. \label{eq:caseB}
\end{equation}
By plugging this expression in Eq.~\eqref{eq:dec3}, we obtain that the following prediction for the lifetime of dark scalars
\begin{equation}
\tau_\phi^{\rm 3-body} \simeq 8.45\times 10^{11}  \left(1+\frac{m_\phi^2}{2 M_R^2}\right)^{-1} \left(\frac{\rm 10^3~GeV}{m_\phi}\right)^2 \left(\frac{106.75}{g^\mathfrak{s}_*}\right)^{3/2}~{\rm sec}\,,
\end{equation}
where, for the sake of simplicity, we have considered the limit $M_R \gg m_\phi \gg m_\chi$. Such values for the lifetime imply that the ordering B case is almost ruled out. Indeed, for a dark scalar mass $10^3~{\rm GeV} \leq m_\phi \leq 10^{12}~{\rm GeV}$ (compatible with the assumptions behind the reported Boltzmann equations), we accordingly have $8.45 \times 10^{11}~{\rm sec} \geq \tau_\phi^{\rm 3-body} \geq 8.45 \times 10^{-7}~{\rm sec}$. This means that the lifetime is too large so that the dark scalars decays into the dark fermions occur after the electroweak breaking or even after the Big Bang Nucleosynthesis. On the other hand, for small dark scalar masses, the lifetime is smaller than the age of the Universe ($\sim 4.35 \times 10^{17}$~sec), scenario that is strongly constrained by cosmological observations and indirect DM searches and in contrast with the two-component DM  assumption. Hence, we find that the ordering type B case can provide an allowed two-component DM scenario only if the dark scalar mass is smaller than the sum of the masses of final particles involved in the three-body decay, i.e. $m_\phi \simeq m_\chi$. In this case, the decays are not kinematically allowed and the dark scalars are stable particles, so providing a viable two-component DM model.

Hence, since the ordering type B is almost ruled out unless $m_\phi \simeq m_\chi$ and the right-handed neutrino coupling is analytically provided by Eq.~\eqref{eq:caseB}, in the next Section we focus only on the numerical results obtained in the ordering type A, i.e. $m_\phi \geq M_R,\,m_\chi$.

\section{Numerical results}

The four free parameters of the model (the three masses of the new particles, $m_\chi$, $m_\phi$ and $M_R$, and the right-handed neutrino portal coupling $y_{\rm DS}$) are constrained by requiring the equality between the observed DM relic abundance~\eqref{eq:omegaOBS}~and the predicted one~\eqref{eq:omegaPREA}. In particular, for a given choice of the seesaw energy scale $M_R$ and the two masses of the dark particles, one can obtain the value for the coupling $y_{\rm DM}$ that provides the correct DM relic abundance by solving Eq.~\eqref{eq:DMyieldDimLess}. 
\begin{figure}[t!]
\begin{center}
\includegraphics[width=0.49\textwidth]{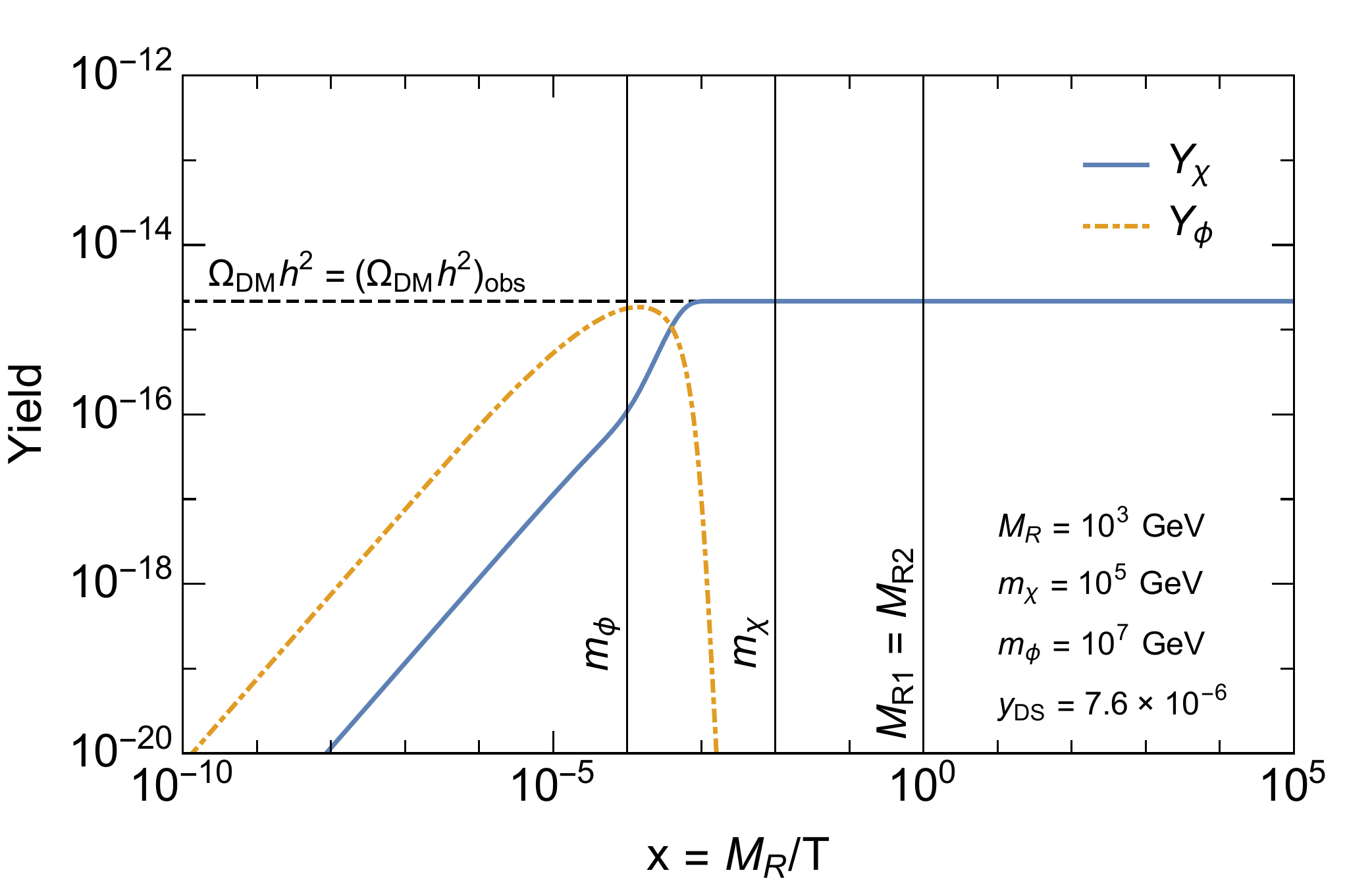}
\includegraphics[width=0.49\textwidth]{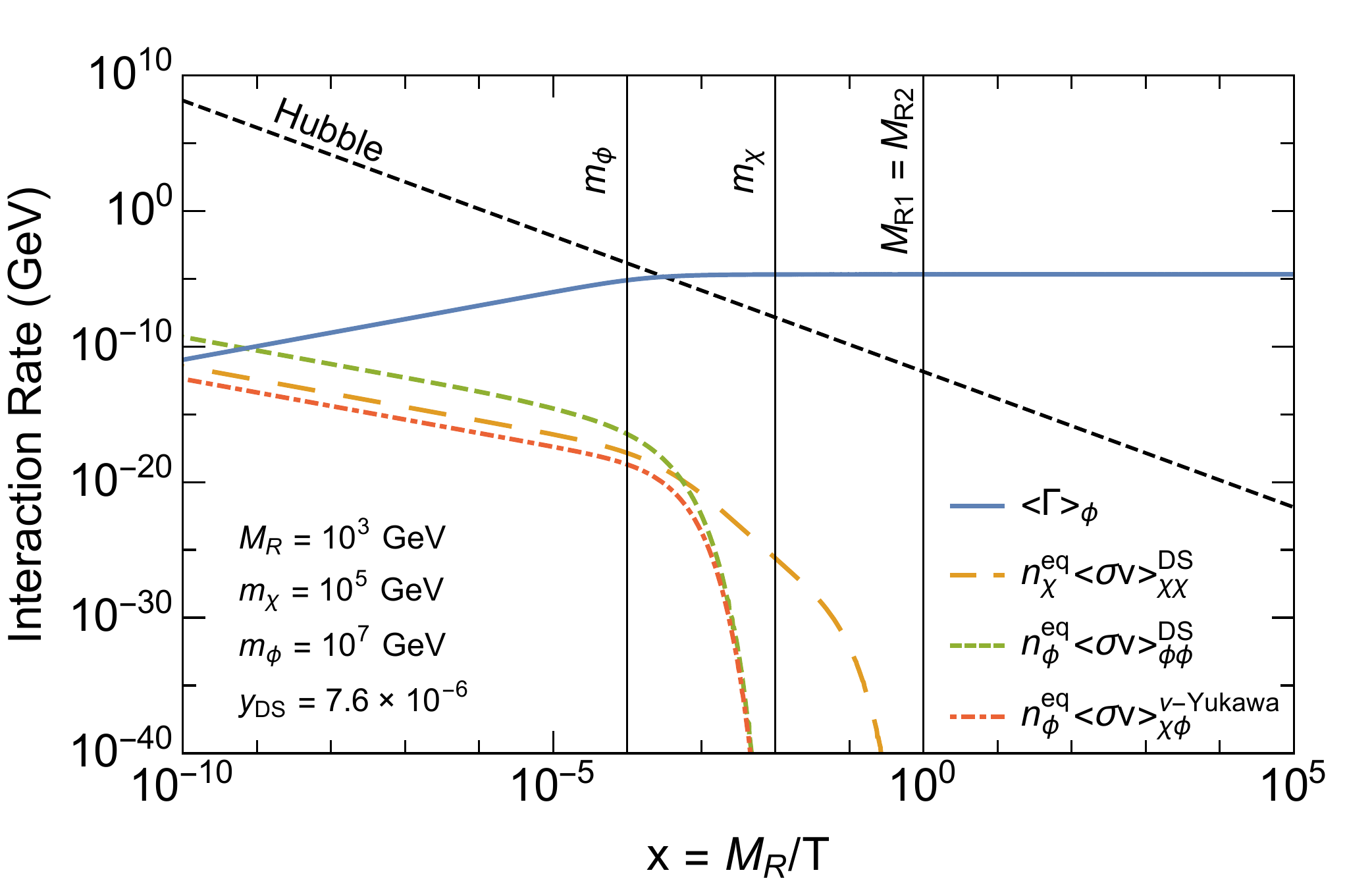}
\caption{\label{fig:benchmark1}Example of DM production through freeze-in mechanism for the benchmark values: $M_R = 10^3$~GeV,  $m_\chi = 10^5$~GeV, $m_\phi = 10^7$~GeV (ordering type A with $M_R \leq m_\chi \leq m_\phi$) and $y_{\rm DS} = 7.5\times 10^{-6}$. {\it Left Panel}: yields of DM particles and dark scalars as a function of the auxiliary variable $x = M_R/T$. {\it Right Panel}: interactions rates of the different processes involved in the Boltzmann equations~\eqref{eq:phi1}~and~\eqref{eq:chi1}.}
\end{center}
\end{figure}

In Fig.~\ref{fig:benchmark1}, we report a benchmark case of ordering A with $M_R = 10^3$~GeV,  $m_\chi = 10^5$~GeV, $m_\phi = 10^7$~GeV and $y_{\rm DS} = 7.5\times 10^{-6}$. As can been seen in the left panel, at high temperatures the DS and $\nu-$Yukawa scattering processes produce particles in the dark sector and the two yields increases as the Universe is cooling. However, such interactions are not strong enough to bring the dark sector in thermal equilibrium. Indeed, the corresponding interaction rates are always smaller than the Hubble parameter, as shown in the right panel of the same figure. At $T=m_\phi$, the yield $\phi$ would freeze-in, but at this temperature its decay rate becomes efficient since $\left<\Gamma_\phi\right> \geq \mathcal{H}$ for $T\lesssim m_\phi$. Hence, the scalar particles are converted into the lighter fermions through the two-body decays that then freeze-in providing the correct DM relic abundance. This conversion can be easily recognised in the left panel by looking to the kink of the $\chi$ yield occurring at $T \simeq m_\phi$. Such a feature suggests that the DM production is mainly driven by the decay of scalar particles that are produced by DS and $\nu-$Yukawa scatterings.
\begin{figure}[t!]
\begin{center}
\includegraphics[width=0.49\textwidth]{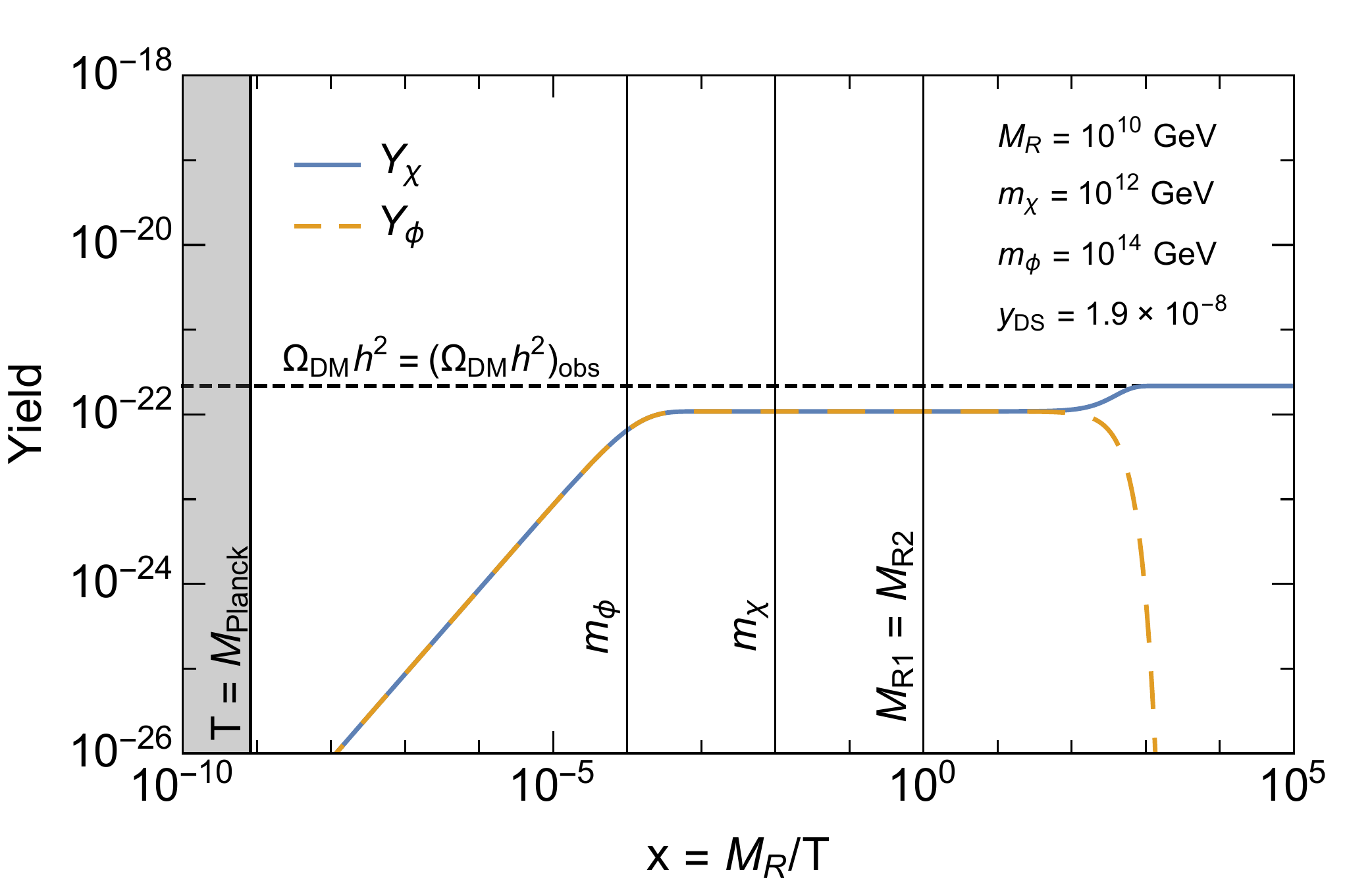}
\includegraphics[width=0.49\textwidth]{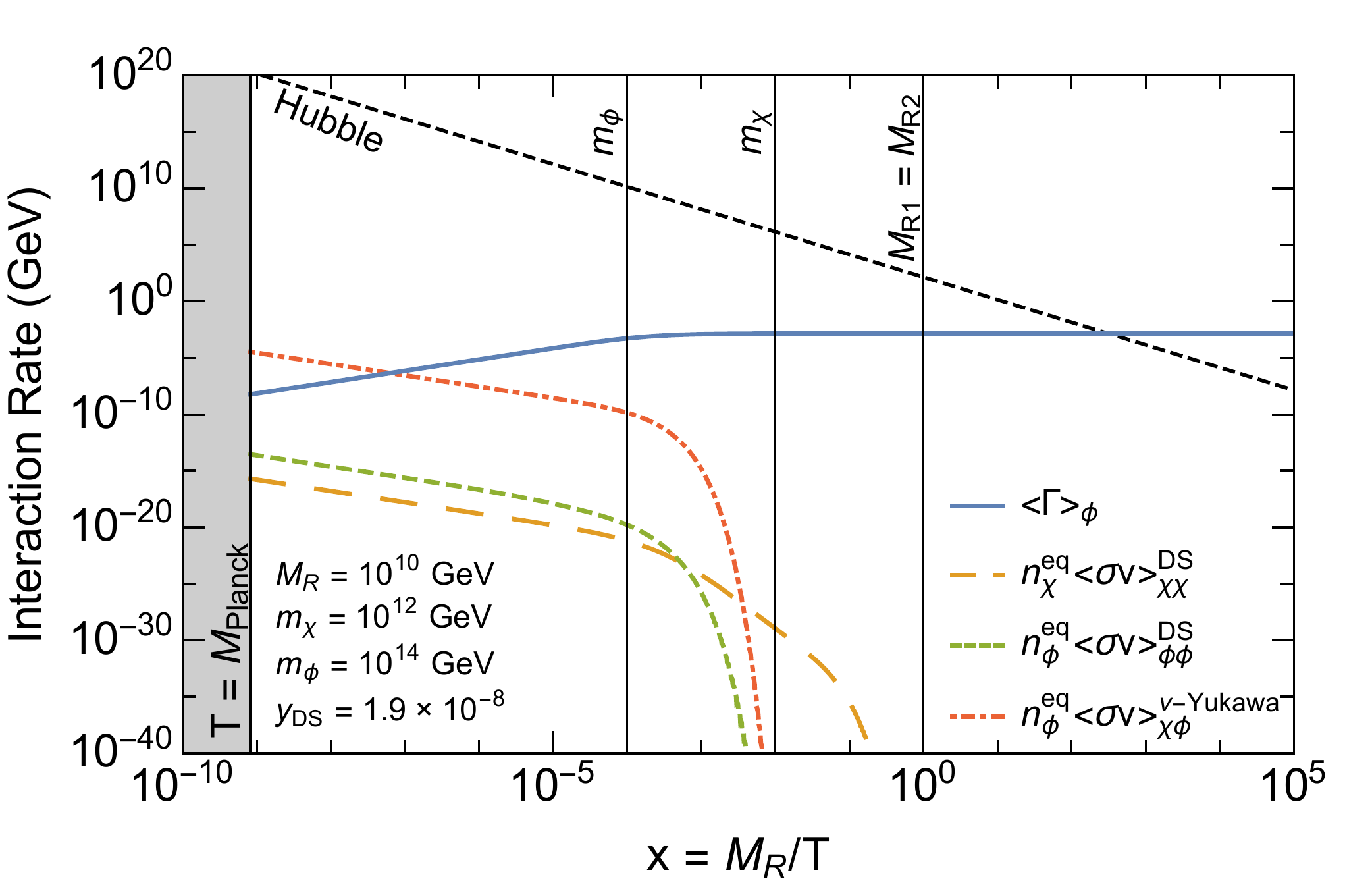}
\caption{\label{fig:benchmark2}Example of DM production through freeze-in mechanism for the benchmark values $M_R = 10^{10}$~GeV,  $m_\chi = 10^{12}$~GeV, $m_\phi = 10^{14}$~GeV (ordering type A with $M_R \leq m_\chi \leq m_\phi$) and $y_{\rm DS} = 1.9\times 10^{-8}$. The description of the plots is the same of Fig.~\ref{fig:benchmark1}.}
\end{center}
\end{figure}

It is worth noticing that, for $M_R=10^3$~GeV, the rates of DS processes are larger than the $\nu-$Yukawa one. This means that the DS processes provide the dominant contribution to the DM production. On the other hand, according to Eq.~\eqref{eq:DMyieldDimLess}, for larger values of the seesaw energy scale $M_R$ the DS processes become less efficient and the DM production starts to be instead driven by $\nu-$Yukawa scatterings. In Fig.~\ref{fig:benchmark2} we show a benchmark case of ordering type A where $\nu-$Yukawa scatterings dominate the DM production. In particular, we consider the values $M_R = 10^{10}$~GeV,  $m_\chi = 10^{12}$~GeV, $m_\phi = 10^{14}$~GeV and $y_{\rm DS} = 1.9\times 10^{-8}$. In this case, both the yields of $\phi$ and $\chi$ particles increase and freeze-in at $T=m_\phi$. This is due to the fact that, as show in the right panel, the dominant contribution to the DM production is provided by the $\nu-$Yukawa processes that simultaneously produce $\phi$ and $\chi$ particles for $T \geq m_\phi$. Then, when $\left<\Gamma_\phi\right> \geq \mathcal{H}$, the scalar particles decay into the dark fermions whose yield just doubles.
\begin{figure}[t!]
\begin{center}
\includegraphics[width=0.49\textwidth]{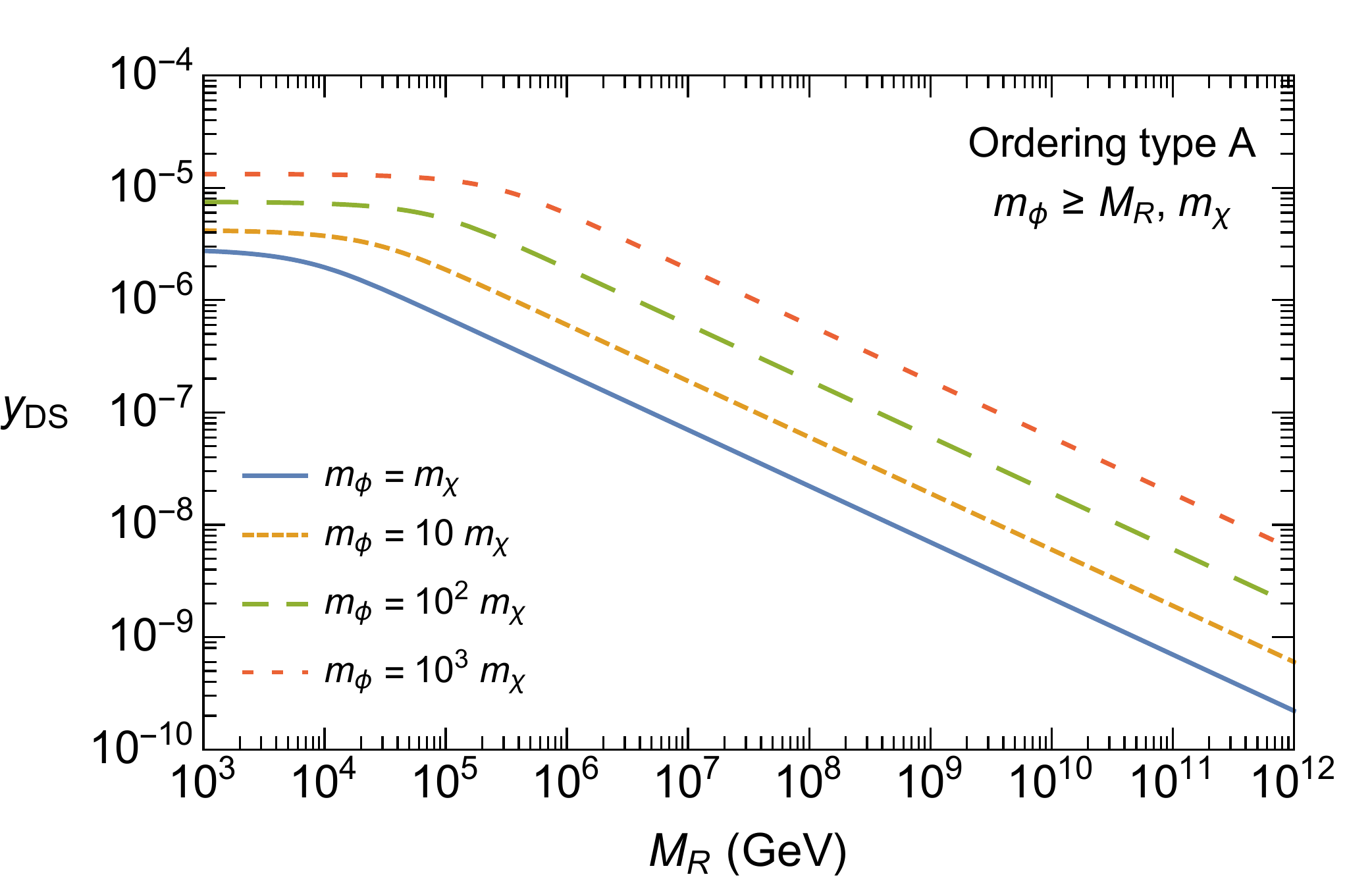}
\includegraphics[width=0.49\textwidth]{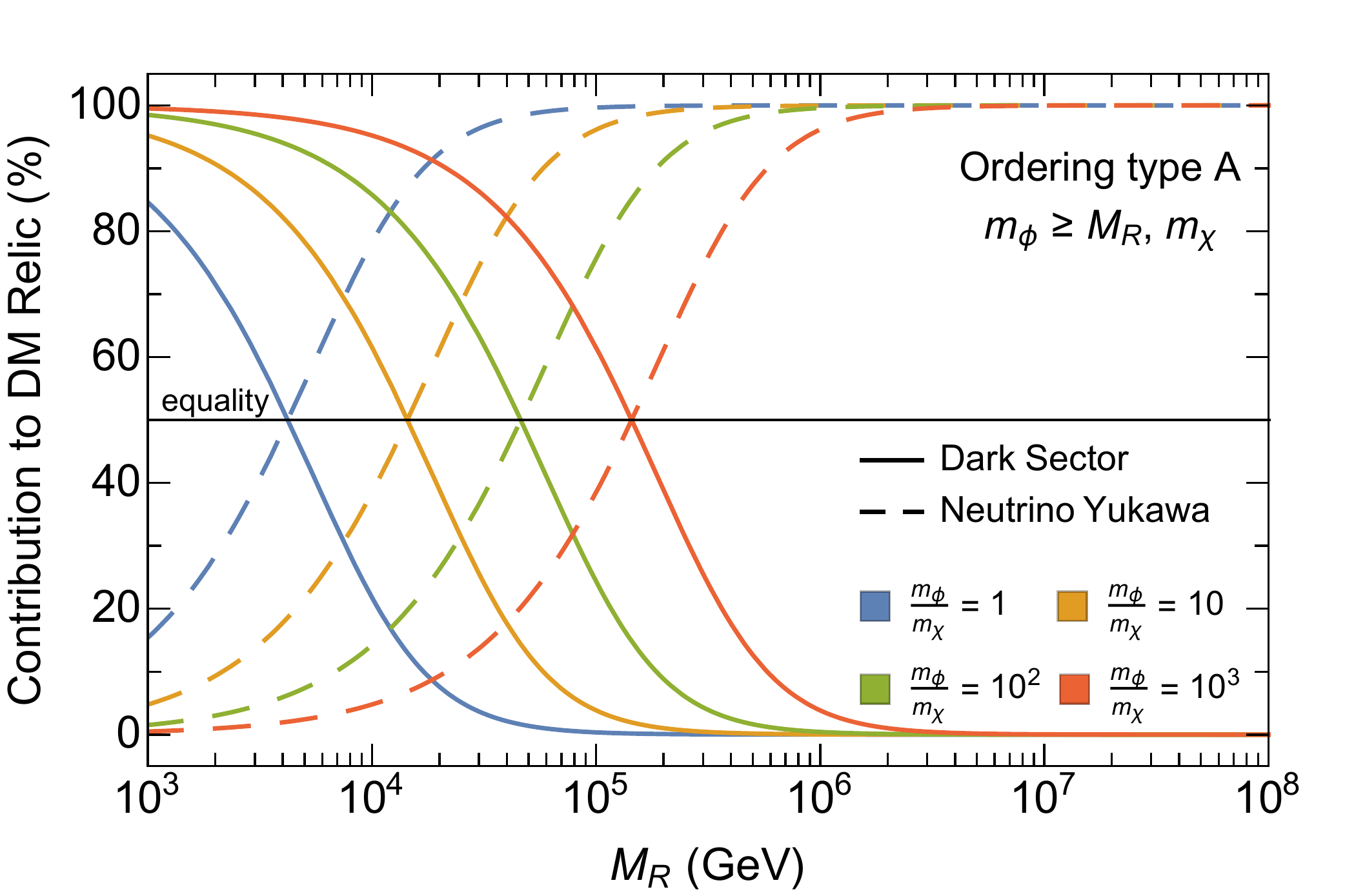}
\caption{\label{fig:RelativeContributions}{\it Left Panel}: right-handed neutrino coupling $y_{\rm DS}$ as a function of the seesaw energy scale $M_R$. {\it Right Panel}: relative contribution of DS (solid lines) and $\nu-$Yukawa (dashed lines) scattering processes to the DM relic abundance. In both panels, the different colours correspond to different values for the ratio $m_\phi/m_\chi$ for ordering type A.}
\end{center}
\end{figure}
\begin{figure}[t!]
\begin{center}
\includegraphics[width=0.65\textwidth]{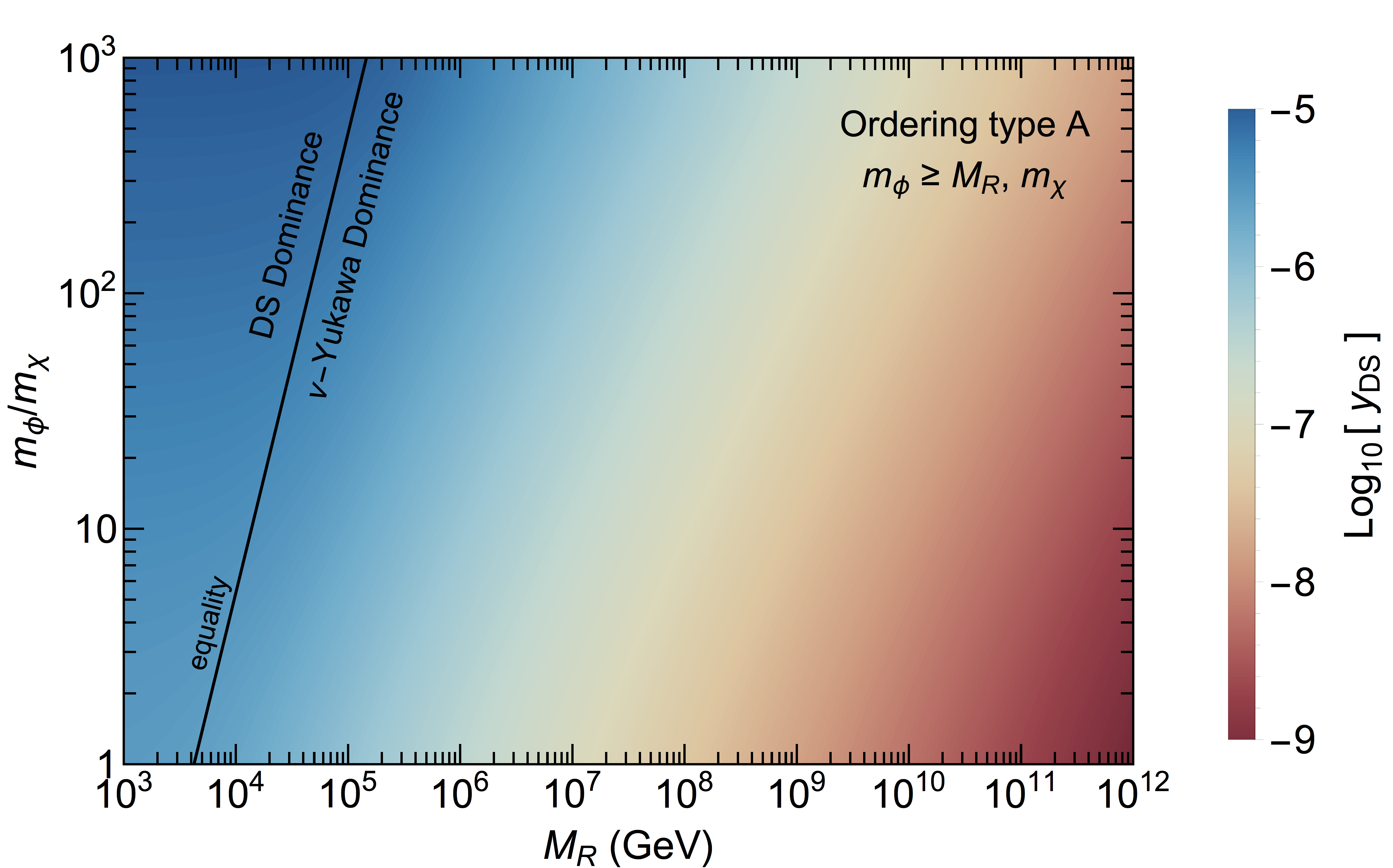}
\caption{\label{fig:NuDMrelation}Right-handed neutrino coupling $y_{\rm DM}$ in the plane $M_R$ -- $m_\phi/m_\chi$ able to reproduce the correct DM relic abundance. The black line highlights the choices of parameters at which the equality of the DS and $\nu-$Yukawa contributions occurs for ordering type A.}
\end{center}
\end{figure}

In Fig.s~\ref{fig:RelativeContributions}~and~\ref{fig:NuDMrelation} we report the main results of the present numerical analysis focusing on the ordering type A ($m_\phi \geq M_R,\,m_\chi$). In particular, in the left panel of Fig.~\ref{fig:RelativeContributions} it is depicted how the right-handed neutrino coupling $y_{\rm DS}$ depends on the right-handed neutrino mass $M_R$. The four different lines correspond to different values for the ratio $m_\phi / m_\chi$, with the blue line representing the case of a two-component DM scenario ($m_\phi = m_\chi$). Hence, we find that the value for the dark coupling required to achieve the correct DM abundance is almost independent on the ratio $m_\chi / M_R$. A slightly different coupling $y_{\rm DS}$ is instead required when $m_\phi \simeq M_R$. However, such a case does not provide a viable DM model since the dark scalars are converted too slowly into dark fermions through the three-body decays as discussed for the ordering type B. For this reason, we do not report here the corresponding results.

Moreover, we note that for small seesaw energy scale the coupling is almost constant, while for larger values of $M_R$ it decreases. Such a change of the behaviour of the coupling $y_{\rm DS}$ occurs at the energy where $\nu-$Yukawa scattering processes start to dominate over the DS ones in the DM production. This can be easily understood by looking to the right panel of Fig.~\ref{fig:RelativeContributions} that shows the relative contribution to the DM relic abundance of the DS (solid lines) and $\nu-$Yukawa (dashed lines) processes. In the plot, the different colours correspond to different values of the ratio $m_\phi / m_\chi$, according to the left panel. In Fig.~\ref{fig:NuDMrelation}, instead, it is displayed the required value for the coupling $y_{\rm DS}$ in the plane $M_R$ -- $m_\phi/m_\chi$. The solid line represents the equality between the contributions to the DM relic abundance of the two different classes of scattering processes. For all the points of the parameter space that provide the correct DM relic abundance with $m_\phi > M_R$, the lifetime of $\phi$ particles is found to be smaller than $10^{-15}$~sec, implying that the scalar decays occur at a temperature higher than the electroweak energy scale. This result confirms the validity of the assumptions behind all the expressions we have used in the present numerical analysis. Finally, we remark that the final DM abundance is found to depend on mass ratio $m_\phi/m_\chi$ only, while it is almost independent on $m_\chi/M_R$. Hence, the results reported in Fig.s~\ref{fig:RelativeContributions}~and~\ref{fig:NuDMrelation} are valid for any choice of masses that satisfies the relations $m_\phi \geq M_R,\,m_\chi$.

\section{Conclusions }

We have proposed a minimal model to simultaneously account for a realistic neutrino spectrum through a type-I seesaw mechanism and a viable dark matter relic density. The model is an extension of the Littlest Seesaw model in which the two right-handed neutrinos of the model are coupled to a $Z_2$-odd dark sector via right-handed neutrino portal couplings. 
In other words we suppose that the production of dark sector particles is achieved dominantly via their couplings to the right-handed neutrinos $N_R$, which are in turn coupled to the thermal bath via their neutrino Yukawa couplings to left-handed neutrinos and Higgs scalar. 

In this model, we have seen that a highly constrained and direct link between dark matter and neutrino physics may be achieved by considering the freeze-in production mechanism of dark matter. In such a framework we have shown that the same neutrino Yukawa couplings which describe neutrino mass and mixing may also play a dominant role in the dark matter production. We have investigated the allowed regions in the parameter space of the scheme that provide the correct neutrino masses and mixing and simultaneously give the correct 
dark matter relic abundance. 

In the above model, we have seen that the results may be classified into two types of cases characterised by whether  the right-handed neutrinos are heavier or lighter than the heavier of the two dark particles. We have seen that the ordering type B with $M_R \geq m_\phi,\, m_\chi$ is almost ruled out unless $m_\phi \simeq m_\chi$, which is the condition that prevents late decays of the dark scalars, and would lead to a viable two-component DM scenario. We have therefore mainly focused on the ordering type A for which $m_\phi \geq M_R, m_\chi$. This is also the most interesting case since then the neutrino Yukawa coupling may play an important role in the production of dark sector particles in the early Universe, and hence in providing the correct relic density.

For type A orderings, we have found that the allowed neutrino portal coupling depends on the right-handed neutrino mass $M_R$ and on the ratio $m_\phi/m_\chi$, while it is almost independent on the ratio $m_\chi/M_R$. Remarkably, for large right-handed neutrino mass, the neutrino Yukawa couplings are fixed via the seesaw mechanism to account neutrino mass and mixing and also by their dominant role in the DM production. The only free parameter is then the common right-handed neutrino mass, which determines the required dark sector right-handed neutrino portal coupling. Such a feature provide a direct link between neutrino physics and dark matter phenomenology, within the framework of the minimally extended Littlest Seesaw model. In certain cases the right-handed neutrino masses may be arbitrarily large, for example in the range $10^{10}-10^{11}$ GeV required for vanilla leptogenesis, with a successful relic density arising from frozen-in dark matter particles with masses around this scale, which we refer to as ``fimpzillas''.

In conclusion, the present paper has made progress in connecting neutrino physics to dark matter in three different ways. Firstly we have proposed a realistic model not only of dark matter but also neutrino mass and mixing, via the Littlest Seesaw model with two right-handed neutrinos, as compared to many neutrino related dark matter models in the literature which only consider a single right-handed neutrino. Secondly, we have considered the freeze-in mechanism for dark matter which is  the first time it has been considered in the literature in the connection with the right-handed neutrino portal.\footnote{Note added: As this paper was being completed, a neutrino portal dark matter model appeared, including regions of parameter space in which the freeze-in mechanism is responsible for dark matter production~\cite{Becker:2018rve}.} Thirdly, we have focused on cases where the same Yukawa couplings which control neutrino mass and mixing also control the relic abundance of dark matter. Within this framework, assuming ordering type A for which $m_\phi \geq M_R, m_\chi$, we have seen that the parameter space is very tightly constrained with the neutrino portal coupling uniquely given by the right-handed neutrino mass $M_R$ for a given ratio  of dark sector masses $m_\phi/m_\chi$. If the right-handed neutrino masses are  in the range $10^{10}-10^{11}$ GeV required for vanilla leptogenesis, this leads to the new idea of superheavy frozen-in dark matter or ``fimpzillas''.

\section*{Acknowledgments}

S.\,F.\,K. acknowledges the STFC Consolidated Grant ST/L000296/1 and the European Union's Horizon 2020 Research and Innovation programme under Marie Sk\l{}odowska-Curie grant agreements Elusives ITN No.\ 674896 and InvisiblesPlus RISE No.\ 690575. M.\,C.\, acknowledges financial support from the STFC Consolidated Grant L000296/1.

\appendix

\section{Amplitudes}
\label{A}

In this Appendix, we report the squared matrix elements for all the scattering processes that are important for the DM production as a function of the corresponding Mandelstam variables $s$, $t$ and $u$. In the following computations, we have used the Feynman rules for Majorana fermions reported in Ref.~\cite{Denner:1992me,Denner:1992vza}. For the DS processes drawn in Fig.~\ref{fig:Feyn1} we have
{\small
\begin{eqnarray}
\overline{\left|\mathcal{M}\right|^2}_{\phi\phi^*\rightarrow n_in_j} & = & - \frac{\left|y_{{\rm DS}i}\, y_{{\rm DS}j}\right|^2}{\left(t-m_\chi^2\right)^2} \left[\left(t-m_\phi^2+M_{Ri}^2\right)\left(t-m_\phi^2+M_{Rj}^2\right)+t\left(s-M_{Ri}^2-M_{Rj}^2\right)\right] \nonumber \\
& & - \frac{\left|y_{{\rm DS}i}\, y_{{\rm DS}j}\right|^2}{\left(u-m_\chi^2\right)^2} \left[\left(u-m_\phi^2+M_{Ri}^2\right)\left(u-m_\phi^2+M_{Rj}^2\right)+u\left(s-M_{Ri}^2-M_{Rj}^2\right)\right]  \\
& & \mp \frac{2 M_{Ri} M_{Rj} \,{\rm Re}\left(y_{{\rm DS}i}\, y_{{\rm DS}j}^*\right)}{\left(t-m_\chi^2\right)\left(u-m_\chi^2\right)} \left[\frac{s+t+u}{2}-2m_\phi^2 \right] \nonumber \,, \\
&& \nonumber \\
\overline{\left|\mathcal{M}\right|^2}_{\chi\overline{\chi}\rightarrow n_in_j} & = &\left|y_{{\rm DS}i}\, y_{{\rm DS}j}\right|^2 \left[\frac{\left(t-m_\chi^2-M_{Ri}^2\right)\left(t-m_\chi^2-M_{Rj}^2\right)}{4\left(t-m_\phi^2\right)^2} +\frac{\left(u-m_\chi^2-M_{Ri}^2\right)\left(u-m_\chi^2-M_{Rj}^2\right)}{4\left(u-m_\phi^2\right)^2} \right] \nonumber \\
& & \mp \frac{M_{Ri} M_{Rj} \,{\rm Re}\left(y_{{\rm DS}i}\, y_{{\rm DS}j}^*\right)\,\left(s-2m_\chi^2\right)}{4\left(t-m_\phi^2\right)\left(u-m_\phi^2\right)} \,,
\end{eqnarray}}
where the sign $-$ ($+$) in the interference term is in case of equal (different) right-handed neutrinos, i.e. $i=j$ ($i \neq j$). On the other hand, neglecting the neutrino and lepton masses, the squared matrix element for the $\nu-$Yukawa scattering processes in Fig.~\ref{fig:Feyn1} take the expressions
{\small
\begin{eqnarray}
\overline{\left|\mathcal{M}\right|^2}_{\phi\overline{\chi}\rightarrow h^0\nu_i} & = & \frac{\left|y_{{\rm DS}1} \left(U_\nu^\dagger Y\right)_{i1}\right|^2}{\left(s-M_{R1}^2\right)^2} \left[\left(s-m_\chi^2-m_\phi^2\right)\left(s-m_{h^0}^2\right)+2 \left(s-M_{R1}^2\right)\left(t-m_\chi^2\right)\right] \nonumber \\
& & +  \frac{\left|y_{{\rm DS}2}  \left(U_\nu^\dagger Y\right)_{i2}\right|^2}{\left(s-M_{R2}^2\right)^2}\left[\left(s-m_\chi^2-m_\phi^2\right)\left(s-m_{h^0}^2\right)+2 \left(s-M_{R2}^2\right)\left(t-m_\chi^2\right)\right] \\
& & + \frac{2\,{\rm Re}\left(y_{{\rm DS}1} \, y_{{\rm DS}2}^*  \left(U_\nu^\dagger Y\right)_{i1} \left(U_\nu^\dagger Y\right)^{*}_{i2}\right)}{\left(s-M_{R1}^2\right)\left(s-M_{R2}^2\right)}\left[\left(s-m_\chi^2-m_\phi^2\right)\left(s-m_{h^0}^2\right)+ 2 \left(s-M_{R1}M_{R2}\right)\left(t-m_\chi^2\right)\right] \nonumber \,, \\
\nonumber
\end{eqnarray}
\begin{eqnarray}
\overline{\left|\mathcal{M}\right|^2}_{\phi\overline{\chi}\rightarrow G^0\nu_i} & = & \frac{\left|y_{{\rm DS}1} \left(U_\nu^\dagger Y\right)_{i1}\right|^2}{\left(s-M_{R1}^2\right)^2} \left[\left(s-m_\chi^2-m_\phi^2\right)\left(s-m_{G^0}^2\right)+2 \left(s-M_{R1}^2\right)\left(t-m_\chi^2\right)\right] \nonumber \\
& & +  \frac{\left|y_{{\rm DS}2}  \left(U_\nu^\dagger Y\right)_{i2}\right|^2}{\left(s-M_{R2}^2\right)^2}\left[\left(s-m_\chi^2-m_\phi^2\right)\left(s-m_{G^0}^2\right)+2 \left(s-M_{R2}^2\right)\left(t-m_\chi^2\right)\right] \\
& & + \frac{2\,{\rm Re}\left(y_{{\rm DS}1} \, y_{{\rm DS}2}^*  \left(U_\nu^\dagger Y\right)_{i1} \left(U_\nu^\dagger Y\right)^{*}_{i2}\right)}{\left(s-M_{R1}^2\right)\left(s-M_{R2}^2\right)}\left[\left(s-m_\chi^2-m_\phi^2\right)\left(s-m_{G^0}^2\right)+ 2 \left(s-M_{R1}M_{R2}\right)\left(t-m_\chi^2\right)\right] \nonumber \,, \\
&& \nonumber \\ 
\overline{\left|\mathcal{M}\right|^2}_{\phi\overline{\chi}\rightarrow G^+\ell^-_i} & = & \frac{\left|y_{{\rm DS}1} Y_{i1}\right|^2}{2 \left(s-M_{R1}^2\right)^2} \left[\left(s-m_\chi^2-m_\phi^2\right)\left(s-m_{G^+}^2\right)+2 \, s \left(t-m_\chi^2\right)\right] \nonumber \\
& & +  \frac{\left|y_{{\rm DS}2}  Y_{i2}\right|^2}{2\left(s-M_{R2}^2\right)^2}\left[\left(s-m_\chi^2-m_\phi^2\right)\left(s-m_{G^+}^2\right)+2 \, s \left(t-m_\chi^2\right)\right] \\
& & + \frac{{\rm Re}\left(y_{{\rm DS}1} \, y_{{\rm DS}2}^*Y_{i1} Y^{*}_{i2}\right)}{\left(s-M_{R1}^2\right)\left(s-M_{R2}^2\right)}\left[\left(s-m_\chi^2-m_\phi^2\right)\left(s-m_{G^+}^2\right)+ 2 \, s \left(t-m_\chi^2\right)\right] \nonumber \,, \\
&& \nonumber \\ 
\overline{\left|\mathcal{M}\right|^2}_{\phi\overline{\chi}\rightarrow G^-\ell^+_i} & = & \left[\frac{\left|y_{{\rm DS}1} Y_{i1}\right|^2 M_{R1}^2}{\left(s-M_{R1}^2\right)^2} +\frac{\left|y_{{\rm DS}2} Y_{i2}\right|^2 M_{R2}^2}{\left(s-M_{R2}^2\right)^2} + \frac{2{\rm Re}\left(y_{{\rm DS}1} \, y_{{\rm DS}2}^*Y_{i1} Y^{*}_{i2}\right) M_{R1}M_{R2}}{\left(s-M_{R1}^2\right)\left(s-M_{R2}^2\right)} \right] \left(t-m_\chi^2\right) \,.
\end{eqnarray}}
By taking the equalities $y_{{\rm DS}1} = y_{{\rm DS}2}$ and $M_{R1} = M_{R2}$, the sum of all the $\nu-$Yukawa scattering processes can be parametrized in terms of the effective squared Yukawa coupling reported in Eq.~\eqref{eq:effyuk}, whose analytical expression is given by
\begin{eqnarray}
\tilde{y}^2_\nu & = & 4 \sum_{i=1}^3\left[\left|\left(U_\nu^\dagger Y\right)_{i1}\right|^2 + \left|\left(U_\nu^\dagger Y\right)_{i2}\right|^2 + 2\,{\rm Re}\left(\left(U_\nu^\dagger Y\right)_{i1} \left(U_\nu^\dagger Y\right)^{*}_{i2}\right) \right] \nonumber \\
& & + \sum_{i=1}^3\left[\left|Y_{i1}\right|^2 + \left|Y_{i2}\right|^2 + 2\,{\rm Re}\left(Y_{i1} Y^{*}_{i2}\right) \right]\,.
\end{eqnarray}
 
\bibliographystyle{JHEP.bst} 
\bibliography{DarkMatter}

\end{document}